\documentclass[structabstract]{aa}  

\usepackage{graphicx}

\usepackage{txfonts}

\begin{document}

\title{Polarization properties of OH masers in AGB and post-AGB stars 
\thanks{Appendices (Table A.1) and (Figure B.1) 
are only available in the electronic edition of the journal at http://www.aanda.org} 
}

\author{P. Wolak \inst{1},
        M. Szymczak  \inst{1}, 
        \and E. G\'erard \inst{2} 
        }

\institute{Toru\'n Centre for Astronomy, Nicolaus Copernicus 
          University, Gagarina 11, 87-100 Toru\'n, Poland 
\and      GEPI, UMR 8111, Observatoire de Paris, 5 place J. Janssen, 92195 Meudon Cedex, France
}

\date{Received 16 May 2011 / Accepted 22 September 2011}

\abstract
{Ground-state OH maser emission from late-type stars is usually polarized and remains 
a powerful probe of the magnetic field structure in the outer regions of circumstellar 
envelopes if observed with high angular and spectral resolutions. Observations
in all four Stokes parameters are quite sparse
 and this is the most thorough, systematic study published to date.}
{
 We aim to
 determine polarization properties of OH masers in 
 an extensive
 sample of stars 
 that show
 copious mass loss and search for candidate objects
 that are
 well-suited for high 
angular resolution studies.}
{Full-polarization observations of the OH 1612 and 1667\,MHz maser transitions 
were carried out for a sample of 117 AGB and post-AGB stars. Several targets were also
observed in the 1665\,MHz line.}
{A complete set of full-polarization spectra together with the basic polarization parameters
are presented. 
{ Polarized features occur in more than 75\% of the sources in the complete
sample and there is no intrinsic difference in the occurrence of polarized emission between 
the three classes of objects of different infrared characteristics. 
The highest fractional polarization occurs for the post-AGB+PN 
and the Mira+SR classes at 1612 and 1667\,MHz, respectively. Differences in the fractional
polarization between the sources at different evolutionary stages appear to be related to 
depolarization 
 caused by
 blending.}
 The alignment of the polarization 
angles at the extreme sides of the shell implies a regular structure of the magnetic field 
of a strength of 0.3-2.3\,mG.}
{Polarized OH maser features are widespread in AGB and post-AGB stars. The relationship between 
the circular and linear fractional polarizations for a 
representative sample
are consistent with the standard models of polarization for
the Zeeman splitting higher than the Doppler line width, whereas the polarized features are the $\sigma$ components.

}
\titlerunning{}
\authorrunning{P. Wolak et al.}
\maketitle

\section{Introduction}

The ground-state OH maser lines at 1612, 1665  and 1667\,MHz are frequently detected in
late-type stars 
that show
 mass loss at high rates. A recent compilation by 
Engels \& Bunzel (\cite{engels08}) gives 2297 circumstellar masers of which 1945 show the classical 
double-peak profile at 1612\,MHz. The polarization information is not available or 
neglected in the interferometric surveys of the Galactic plane 
(Sevenster et al.\,\cite{sevenster97a, sevenster97b, sevenster01})
in numerous single-dish studies of IRAS colour-selected objects (e.g. Eder et al.\,\cite{eder88}; 
Le Squeren et al.\,\cite{lesqueren91}; te Lintel Hekkert et al.\,\cite{telintel91}) 
and in earlier blind surveys (e.g. Bowers\,\cite{bowers78}; Baud et al.\,\cite{baud79a, baud79b}).  
Earlier studies (e.g. Wilson et al.\,\cite{wilson70}; Wilson \& Barrett\,\cite{wilson72}) 
suggested that the 1612\,MHz circumstellar emission is essentially unpolarized but the observations 
with a spectral resolution better than 0.1\,km\,s$^{-1}$ revealed several circularly polarized 
features (Cohen et al.\,\cite{cohen87}; Zell \& Fix\,\cite{zell91}). The OH emission in the 1665 
and 1667\,MHz lines shows a substantial degree of circular polarization up to 50$-$80\% 
(Wilson et al.\,\cite{wilsonetal72}; Claussen \& Fix\,\cite{claussen82}) but linear polarization 
occurs in only $\sim$10\% of sources (Olnon et al.\,\cite{olnon80}). Highly polarized OH maser 
features are observed in flaring Miras (Reid et al.\,\cite{reid77}; Fix\,\cite{fix79}; 
Etoka \& Le Squeren\,\cite{etoka97}), hypergiant IRC+10420 (Benson et al.\,\cite{benson79}) and 
post-AGB star OH17.7-2.0 (Szymczak \& G\'erard\,\cite{szymczak05}).
There is a general consensus that the polarization measurements of OH maser emission can be properly 
explained with the classic model of Goldreich et al. (\cite{goldreich73}) in which the magnetic 
field plays a key role (e.g. Gray \& Field\,\cite{gray95}; Elitzur\,\cite{elitzur96}). 

The presence of circular polarization is itself evidence for Zeeman splitting 
(Deguchi \& Watson\,\cite{deguchi86}) and, despite the difficulties in identifying classical Zeeman 
patterns, provides a direct estimate of the strength and direction of the magnetic field. 
From observations of the linear polarization 
we can draw
 conclusions about the orientation
of the circumstellar magnetic field.

For some OH stellar sources, high-sensitivity observations with $\sim$0.1\,km\,s$^{-1}$ spectral 
resolution have shown an amount of circularly and linearly polarized features large enough 
to measure the strength of circumstellar magnetic fields and to map their structure with high 
accuracy using the interferometric technique (e.g. Chapman \& Cohen\,\cite{chapman86}; 
Szymczak et al.\,\cite{szymczak98, szymczak01}; Bains et al.\,\cite{bains03}; 
Amiri et al.\,\cite{amiri10}). These case studies revealed the presence of highly ordered magnetic 
fields of a strength of a few mG at distances of a few thousands stellar radii.
Full-polarization measurements of OH circumstellar masers with high sensitivity and spectral 
resolution are scarce. A small sample of Miras (Claussen \& Fix\,\cite{claussen82}) and 
a
moderate 
sample of proto-planetary nebula candidates (Szymczak \& G\'erard\,\cite{szymczak04}) only were 
studied up to date and no comprehensive picture of the polarization properties of circumstellar 
OH masers was obtained. We hereby extend those studies for a sample containing oxygen-rich objects 
that
 evenly populate the whole evolutionary sequence in the IRAS two-colour diagram 
(van der Veen \& Habing\,\cite{vanderveen88}). There is evidence that the IRAS colours increase 
through the sequence from the semi-regular variables (SRs), long period variables of Mira types, 
OH/IR objects, proto-planetary nebulae (PPN) and planetary nebulae (PN). 
The colours probably reflect the evolution of the circumstellar shell 
from optically thin to optically thick and the location of an object in the diagram could be also 
related to the age, initial mass and possibly metallicity (e.g. Bedijn\,\cite{bedijn87}; 
van der Veen \& Habing\,\cite{vanderveen88}; Likkel\,\cite{likkel89}). The observations of a 
representative sample reported here attempt to examine whether the OH maser polarization depends 
on the properties of the central star and its envelope or whether it is caused by 
propagation effects.

\section{Data}

\subsection{The sample}
 The initial source list consisted of 152 late-type stars at $\delta \ga -39$\degr\, 
of which 147 objects 
were taken
 from Benson et al.'s catalogue (\cite{benson90}), 
the rest 
from te Lintel Hekkert et al.'s survey (\cite{telintel91}). 
Several sources were discarded after the initial test
observations because their total peak fluxes were lower than 0.15\,Jy at both frequencies. The final sample
consisted of 117 stellar OH sources that clearly exhibit either the 1612\,MHz satellite line 
or/and the 1665/1667\,MHz main lines.
Targets with different infrared characteristics were chosen  
to probe the whole evolutionary sequence of oxygen-rich objects in the IRAS two-colour diagram 
(van der Veen \& Habing\,\cite{vanderveen88}). 
The location of an object in this diagram depends on the circumstellar envelope thickness, age and 
initial mass. It is assumed that the visible semi-regular and Mira variables with relatively thin envelopes 
evolve towards OH/IR objects with thick envelopes, and when their pulsations cease, they leave the tip of 
the asymptotic giant branch and become post-AGB objects and eventually young planetary nebulae
(Habing\,\cite{habing96}, for review). 
Alternatively, OH/IR stars have higher progenitor masses than Miras (Likkel\,\cite{likkel89})
and may come from an intrinsically more massive population (Whitelock et al. \cite{whitelock94}).
 These categories of objects differ in pulsational and mass-loss
properties and can be accompanied by OH maser emission.

\begin{figure}
   \resizebox{\hsize}{!}{\includegraphics[angle=-90]{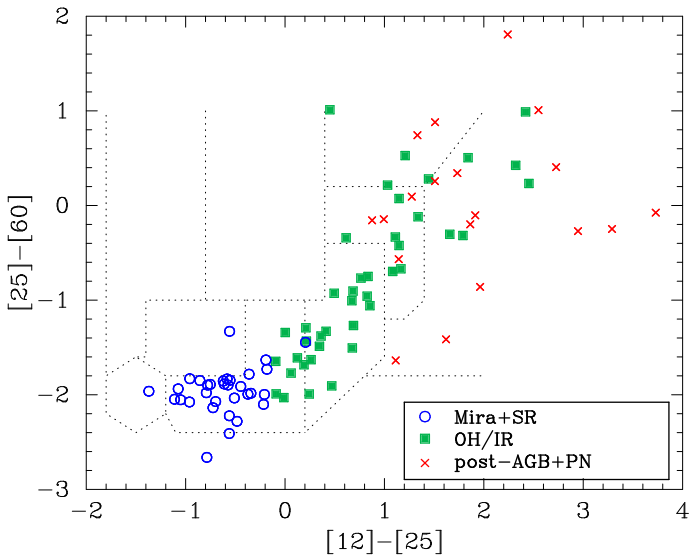}}
   \caption{IRAS colour-colour diagram for the sample studied. The regions defined by 
    van der Veen \& Habing (\cite{vanderveen88}) for different types of objects are marked by dotted lines. 
   The three classes of objects discussed in the text can be distinguished by the symbols.} 
   \label{fig1}
\end{figure}

There are 36 SRs and Miras, 57 OH/IR objects and 24 post-AGB and PN candidates 
in the sample (Table A.1).
We are concerned here only with these three groups of objects. 
Note that the latter group was  
included in a sample studied by Szymczak \& G\'erard (\cite{szymczak04}) but the spectra used there
were generally taken at other epochs.
One hundred and five out of 117 targets have good flux quality at the 12, 25 and 60$\mu$m IRAS bands, and 
we show their two-colour diagram in Figure \ref{fig1}. The colour is defined as 
$[\lambda_i] -[\lambda_j]= -2.5$log(S($\lambda_i$)/S($\lambda_j$)), 
where S($\lambda_i$) is the uncorrected
flux in the IRAS $\lambda_i$ band. The colours of OH/IR and post-AGB objects overlap, partly because of
inaccurate classification criteria of the transition objects between AGB and post-AGB phases.
 Figure \ref{fig1} shows that our targets sample all types of oxygen-rich
stars with a circumstellar envelope associated with OH maser emission. 
The sample is therefore 
appropriate to properly address whether the polarization parameters of OH masers depend on 
the characteristics of the central stars/envelopes or are dominated by propagation effects.

 A comparison of the 1612 MHz peak flux density for 105 
detected sources in the sample with that for 808 objects in the te 
Lintel Hekkert et al. (\cite{telintel91}) catalogue indicates that our 
sample probes the brightest sources; the median values 
are 8.32 and 0.86 Jy, respectively (Table \ref{samples-properties}). We note, however, that 
the older observations may underestimate the peak flux because of frequency 
dilution.

\subsection{The observations}

The data were obtained with the Nan\c{c}ay Radio Telescope (NRT) during several runs between 
February 2002 and December 2008. This transit instrument is the equivalent of a dish of 93\,m 
in diameter and the half-power beam-width is $3\farcm5$(RA)$\times 19\arcmin$(Dec) at 1.6\,GHz.  
At Dec = 0\degr, the beam efficiency is 0.65 and the point-source efficiency is 1.4\,K\,Jy$^{-1}$, 
the system temperature is about 35\,K. More technical details on the radio telescope 
and its upgrade are given in van Driel et al. (\cite{vandriel96}).

Initially, we observed 152 sources in the 1612.231 and 1667.359\,MHz OH
lines, 38 of which were also observed at 1665.402\,MHz 
(including 31 sources that were not detected at 1612.231\,MHz).
We used a correlator configuration providing eight banks of 1024
channels. 
To cover all OH maser profiles easily, we used 
two spectral resolutions, $\sim$0.14\,km\,s$^{-1}$ for about two thirds 
of the targets and $\sim$0.07\,km\,s$^{-1}$ for the rest.

Two transitions were simultaneously 
observed and we registered the two orthogonal linear polarizations and two opposite 
circular polarizations for each transition. 
The system directly provided three of the four Stokes parameters, namely I, Q and V, 
while the fourth parameter U was extracted by a horn rotation of 45\degr.
The linearly polarized flux density, $p=\sqrt{Q^2 +U^2}$, fractional linear polarization,
$m_{\mathrm L}=p/I$, fractional circular polarization, $m_{\mathrm C}=V/I$ and polarization position angle, 
$\chi=0.5\times $tan$^{-1}(U/Q)$, were derived from the Stokes parameters. The methods of observation, 
polarization calibrations, data reduction and estimation of uncertainties on the Stokes parameters 
were the same as described in Szymczak \& G\'erard (\cite{szymczak04}, Sect. 2). 
 Briefly, we observed in frequency
switching mode, using a noise diode to measure the gain of each
polarization bank at the start of each scan.
The absolute and relative accuracies 
of gain were $\sim$5\% and $\sim$1\%, respectively. The instrumental polarization was regularly 
checked with observations of W12 and W3OH and the absolute flux density scale was accurate to 
within 7$-$8\%. 
 The error in the polarized intensity caused by the polarization leakage 
between the orthogonal feeds was about 2\%.
The typical integration time for each horn position was about 20\,min. This resulted 
in a rms noise level in the Stokes $I$ of about 35\,mJy for 0.14\,km\,s$^{-1}$ spectral resolution. 
The median
3$\sigma$ rms noise levels of the Stokes $Q$ and $U$ spectra were 0.15\,Jy.
In those features that showed polarization, the polarization was only
considered believable
 if either or both of the linearly or circularly
polarized fluxes were higher than 0.15 Jy.

The data for our sample have been collected over nearly seven years. Each target was observed at least 
three times on an irregular basis but there are several sources that were observed over a 3$-$7 years
period so that the light curves could be easily obtained. Here we only report data for the epoch
during which the maximum flux density occurred. Results of the whole long-term observations will be published 
in a separate paper.

\section{Results and analysis}

 Table A.1 contains a list of the 117 detected 
sources and their polarization parameters. 
Full-polarization spectra are shown in Figure B.1.

\subsection{Effect of spectral resolution}

 We tested whether the fractional polarization decreases as the spectral 
resolution is degraded by observing a few
targets at both spectral resolutions.
 For the two sources OH17.7$-$2.0 
and OH138.0$+$7.2 the maximum values of $m_{\mathrm C}$ and $m_{\mathrm L}$ at 1612 and 1667\,MHz decreased 
by less than 0.2\% and 0.5\%, respectively, when the resolution was lowered from $\sim$0.07 to 
$\sim$0.14\,km\,s$^{-1}$. Additionally, the 1612 and 1667\,MHz spectra of $\sim$0.07\,km\,s$^{-1}$ 
resolution with detected polarized features were smoothed to a resolution of $\sim$0.14\,km\,s$^{-1}$.
The average and median values of the fractional polarizations for the strongest features before and 
after the smoothing are given in Table \ref{two-resolution}. No significant difference in the fractional polarizations 
is found for both observed lines. These two tests clearly indicate that our data taken with the two different 
spectral resolutions can be treated together in the subsequent analysis.

\setcounter{figure}{1}

\begin{table}

\caption
{Comparison of
mean and median peak flux densities in our sample and that of 
te Lintel Hekkert et al. (\cite{telintel91}).}
\label{samples-properties}

\begin{tabular} {l c c c c}

 \hline  

 sample            & number       & frequency  & mean(SD$^a$)  &  median \\ 

                   &  of sources  &     (MHz)  & (Jy)  &  (Jy)   \\

 \hline

                    &       &  &   &    \\ 

 this paper         &   105    & 1612  & 31.70 (8.12) &  8.32   \\

                    &   85     & 1667  & 3.59 (0.74) &  1.21   \\

                    &       &  &   &    \\

 te Lintel Hekkert  &   808     & 1612  & 4.37 (0.84)  &  0.86  \\   

 \hline

 \end{tabular}
\tablefoot{$^a$ standard deviation}
\end{table}

\begin{table*}
\caption{Average and median values of the fractional circular ($m_{\mathrm C}$) and linear ($m_{\mathrm L}$) polarization 
         for the strongest features at the two spectral resolutions.}
\label{two-resolution}
\begin{tabular}{l c c c c c c }

\hline\hline

Line &   Fractional &  Number     & Mean(SD)$^o$& Median$^o$ &  Mean(SD)$^s$& Median$^s$ \\ 

(MHz)&   polzn. & of sources     & (\%)    &  (\%)  &   (\%)   &  (\%)  \\

\hline

1612 &   $|m_{\mathrm C}|$ & 11 & 7.5(1.8)& 5.8 & 8.2(2.3) & 5.5 \\         

     &   $m_{\mathrm L}$ & 12 & 4.5(0.6)& 4.0 & 4.2(0.6) & 3.9 \\

1667 &   $|m_{\mathrm C}|$ & 12 & 27.0(9.4)& 10.1 & 27.0(9.0) & 10.2 \\

     &   $m_{\mathrm L}$ &  6 & 7.1(2.6)& 4.2 & 7.5(2.6) & 5.6 \\

\hline

\end{tabular}

Spectral resolution: $^o \sim$0.07\,km\,s$^{-1}$; $^s \sim$0.14\,km\,s$^{-1}$

\end{table*}

\begin{table}

\caption{Number of sources with polarized features at the OH lines.}

\label{n_polzn_feat}

\begin{tabular}{l c c c c}

\hline\hline

Line  & Number of & \multicolumn{3}{c}{Number of sources with features} \\

\cline{3-5}

(MHz) & detection & elliptical & circular & linear\\

\hline

1612 only &    32          &    10  &   4  &   3 \\

1667 only &    12          &         3   &       7   &       0  \\

1612/{\it 1667}&    73     &    52  &  16  &   7 \\

1667/{\it 1612}&    73     &        15   &   8  &   4 \\

\cline{1-5} 

\multicolumn{2}{c}{total} &     &    & \\

\hline

1612  &   105     &      58  &    20 &   10 \\

1667  &    85     &          18   &    15 &    4 \\

\hline

\end{tabular}

\end{table}

\subsection{Occurrence of polarized emission}

The detection of polarization is limited by the noise level of 0.15\,Jy in
any Stokes parameter and the minimum believable fractional polarization of 2\%.
Thus, our sample is complete for objects that are higher than 2\% polarized
and whose total intensity exceeds 7.5\,Jy. The weaker sources only have
detectable polarization if they are more strongly polarized.
Table \ref{n_polzn_feat} summarizes the detection statistics for elliptical, circular and linear
polarizations towards the sources divided into three groups depending on the occurrence
of the 1612\,MHz or/and 1667\,MHz lines. Column 2 gives the number of detected sources
and columns 3-5 list the number of sources with features of given types of polarization.
The emission at 1612 and 1667\,MHz was detected in 105 and 85 targets, respectively.
 Polarized features were detected in
 75\% (79/105) and  46\% (39/85) of the sources at 
1612\,MHz and 1667\,MHz, respectively. This is because the 1612\,MHz
emission is generally stronger than the 1667\,MHz emission.
The study indicates that elliptical polarization dominates at 1612\,MHz. There is
also a high detection rate of linearly polarized emission 
( 58\% at 1612\,MHz and  26\% at 1667\,MHz) in the sample, never reported previously
(Cohen\,\cite{cohen89}, for review).  Surprisingly, we detected  14 sources 
with linearly polarized features without apparent circular polarization (see Sect.3.5).

\subsection{Fractional polarization}

The polarization properties of maser spectra are commonly determined relative
to the peak flux density of Stokes $I$ parameter. However, we found that the
velocities of the Stokes $I$ peak do not coincide with those of the Stokes $V$ 
and $p$ peaks for 62\% and 65\% of the sources, respectively.

The median velocity differences between the peaks of Stokes $I$ and Stokes
$V$ and $p$ are 0.14 and 0.27\,km\,s$^{-1}$, respectively. 
Generally, the most strongly polarized features coincide with the Stokes
$I$ peaks within 2-3 spectral channels.
However, for a few sources these velocity differences are as large as
5.2\,km\,s$^{-1}$ in IRAS19114$+$0002, 1.7\,km\,s$^{-1}$ in 
OH12.8$-$1.9 and 0.9\,km\,s$^{-1}$ in RR Aql, then the $V$ and $p$ peaks 
are distinct features.
 Velocity separations could be caused by Zeeman splitting (see Sect. 3.6).
 Blending of differently
polarized components along the line of sight would also cause
depolarization relative to the total intensity. In the following
sections, we perform separate statistical analyses for the
fractional polarization at the Stokes $I$, Stokes $V$ and $p$ peaks. We
compare the results for the three classes of objects defined in Table A.1.

\subsubsection{Measurements relative to the Stokes $I$ peak}

\begin{figure*}
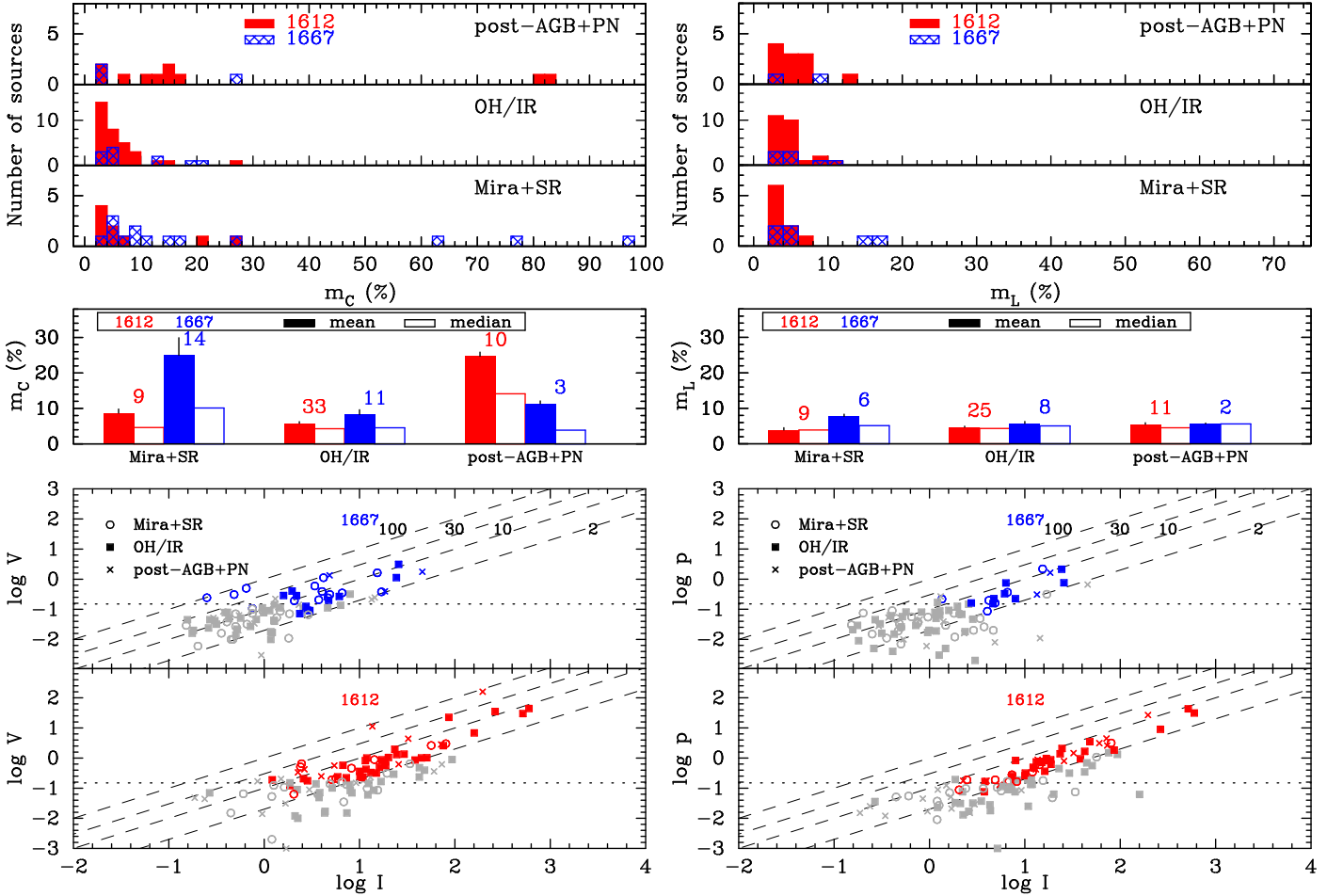


\includegraphics[scale=0.475]{peaks_2_1AL.ps}
\includegraphics[scale=0.475]{peaks_2_2AL.ps}

\caption{Polarization properties of OH emission in three classes of late-type stars. 
 The flux density measurements are made at the Stokes I peak.
           {\it Top:} Histograms of circular ($m_{\mathrm C}$) and linear ($m_{\mathrm L}$)
           fractional polarizations for each class of objects at both OH lines.
           {\it Middle:} Mean and median values of the circular and linear fractional 
           polarizations for three classes of objects. The number of objects for each class and 
           OH line is given.
           {\it Bottom :} Circularly polarized peak flux density (Stokes $|V|$)
           ({\it left}) and linearly polarized peak flux density (parameter $p$)({\it right}) versus 
           the total peak flux density (Stokes $I$) for the two OH lines. The coloured (online) 
           symbols represent classes of objects while the grey symbols mark the sources with 
           polarized emission lower than the individual detection level (3$\sigma$). 
            The dotted lines show the mean upper limit ~0.15 Jy.
Some objects have much lower individual upper limits because they were
observed longer. The dashed lines labelled 2, 10, 30 and 
           100 indicate the percentage of polarization.}
   \label{polznprop}

\end{figure*}

\begin{table*}
\caption{Mean and median values of fractional polarization determined  at the velocity of the Stokes $I$ peak 
flux density.\label{frac_polzn_Ip}}

\begin{tabular} {c c c c c c c c }
  \hline\hline 
 Class & Line  & number & \multicolumn{2}{c}{$|m_{\mathrm C}|$(\%)} & number & \multicolumn{2}{c}{$m_{\mathrm L}$(\%)}\\ 
 \cline{4-5}
 \cline{7-8}
       & (MHz) & of sources & mean (SD) & median  & of sources    & mean (SD)  & median \\ 
\hline

Mira+SR     & 1612 &  9  &  8.70(1.14)  &  4.62 &  9 &  3.89(0.66) &  3.89\\

            & 1667 &  14 &  24.98(4.95) &  10.13 &  6 &  7.70(0.66) &  5.14\\

OH/IR       & 1612 &  33 &  5.81(0.48) &  4.28  &  25 &  4.70(0.29) &  4.34\\

            & 1667 &  11 &  8.28(1.36) &  4.55  &  8 &  5.71(0.59) &  5.05\\
 
post-AGB+PN & 1612 &  10 &  24.71(1.17) &  14.14 &  11 &  5.50(0.44) &  4.52\\

            & 1667 &  3  &  11.30(0.82) &  3.87  &  2  &  5.60(0.24) &  5.60\\


\hline

\end{tabular}

\end{table*}

Figure \ref{polznprop} summarizes the polarization properties of OH emission for the three classes of objects
in the sample, and Table \ref{frac_polzn_Ip} gives the mean and median values of circular and linear
fractional polarization.
The fractional circular polarization in the OH sources is commonly lower than 20\% and is as high as 
60$-$100\% in a few objects. In contrast, the fractional linear polarization is usually lower than 10\% 
 and reaches a slightly higher value in a few sources.
 The differences in the fractional polarization distributions between the three
classes and the two lines were investigated using a Kolmogorov-Smirnov test. 
In the post-AGB$+$PN objects the mean value of $m_{\mathrm C}$ at 1612\,MHz is
higher ($p<0.01$) than that for the other two classes. 
The mean degree of circular polarization at 1667\,MHz of the Mira$+$SR
objects is higher ($p<0.01$) than that of the OH/IR and post-AGB$+$PN objects.
There is no significant difference in the distribution of $m_{\mathrm L}$ between the three classes
of objects.
 There are 10 Mira+SR stars, 21 OH/IR objects and 8 post-AGB+PN sources with $I_p>$10\,Jy 
without detectable polarization (Fig. \ref{polznprop}).

\subsubsection{Measurements relative to the Stokes $V$ and $p$}

\begin{table*}

\caption{Mean and median values of fractional circular and linear polarization determined  at the velocities of 
the Stokes $V$ and $p$ peak flux densities, respectively.\label{frac_polzn_Vppp}}

\begin{tabular}{c c c c c c c c}

\hline\hline

 Class & Line  & number & \multicolumn{2}{c}{$|m_{\mathrm C}|$(\%)} & number & \multicolumn{2}{c}{$m_{\mathrm L}$(\%)}\\ 

 \cline{4-5}

 \cline{7-8}

       & (MHz) & of sources & mean (SD) & median  & of sources    & mean (SD)  & median \\ 

\hline

Mira+SR       & 1612 &  14 &  12.62(1.68) &  5.57  & 12   & 7.13(1.15)  & 4.07\\

              & 1667 & 18 & 39.31(8.76) & 26.53 & 8    & 11.34(0.93) & 8.56\\

OH/IR         & 1612 &  38 &  7.37(0.68)  &  6.13  &  37   &  5.46(0.40)  &  4.75\\

              & 1667 & 12 & 11.54(1.71) &  8.70 &  9    &  11.32(1.41) &  8.77\\

post-AGB+PN   & 1612 & 18 & 23.41(3.47) & 13.47 & 11   & 6.79(0.44)  & 5.19\\

              & 1667 &  6  &  24.51(3.42) &  26.28 &  4    &  4.23(0.24)  &  2.96\\


\hline

\end{tabular}

\end{table*}

\begin{figure}   

   \includegraphics[angle=0,width=0.5\textwidth]{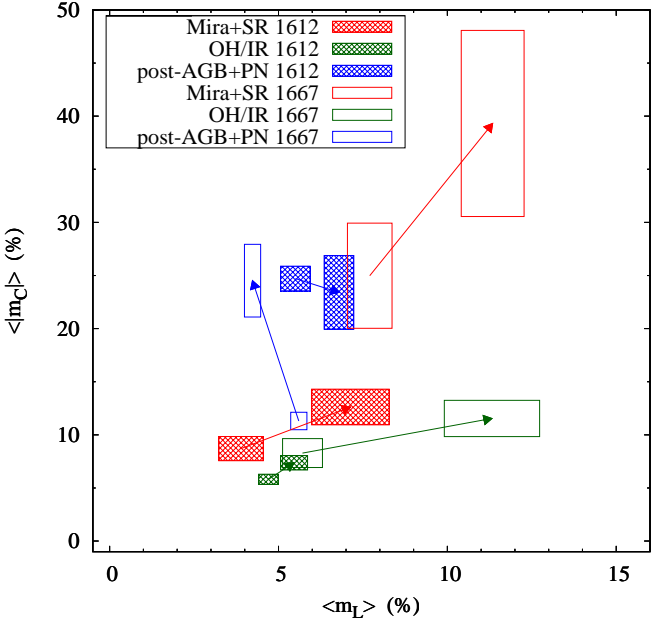}

  \caption{Comparison of the mean fractional polarizations obtained with two different methods
          (Tables \ref{frac_polzn_Ip} and \ref{frac_polzn_Vppp}). The mean values of $m_{\mathrm L}$ and $m_{\mathrm C}$ 
          correspond to the rectangle centres and the standard deviations are marked by the borders. 
          The origins of the arrows mark the mean fractional polarizations estimated at 
          the Stokes $I$ peaks, whereas the arrowheads mark the mean fractional polarizations
          estimated at the Stokes $V$ and $p$ peaks.\label{mean_med_boxes}}
\end{figure}

The velocity offset of the 
Stokes $V$ and $p$ peaks from the Stokes $I$ peaks, as noted in 
Sect. 3.3, is most pronounced for the Mira+SR stars where it is 
 observed in 70--75\% of the objects.
The result of the analysis of the fractional polarization based on the measurements 
concerning the peak fluxes in the $V$ and $p$ parameters is summarized in 
Table \ref{frac_polzn_Vppp}. 
 Generally, the mean and median values of $m_{\mathrm C}$ and 
$m_{\mathrm L}$ are 2$-$3 times
 higher than the values measured at the Stokes $I$
 peak (Table \ref{frac_polzn_Ip}). The increase in $m_{\mathrm C}$ is greater at 1667
 than at 1612 MHz. 
There is a remarkable increase of both $|m_{\mathrm C}|$ 
and $m_{\mathrm L}$ for the 1667\,MHz Mira$+$SR objects (Fig. \ref{mean_med_boxes}).
The trends described in Sect. 3.3.1 remain valid. We also measured the dispersion of
$m_{\mathrm C}$ and $m_{\mathrm L}$ in consecutive spectral channels with polarized signal.
It appears that at 1612\,MHz the dispersion of $m_{\mathrm C}$ for the Miras+SR and post-AGB+PN
stars is nearly twice higher than in the OH/IR objects. The Mira+SR variables 
also show the highest dispersion of $m_{\mathrm C}$ at 1667\,MHz. 
The present
analysis clearly shows that for the majority of sources the most highly
polarized features are weaker than the line peaks. 
In the following analysis of the polarization properties, the measurements relative to the Stokes 
$V$ and $p$ are used because they are the best estimates of the maximum values of the polarization
parameters in the sources (Tables \ref{frac_polzn_Ip} and \ref{frac_polzn_Vppp}).

\subsubsection{Ratios of the integrated flux densities}

\begin{figure}   
   \includegraphics[angle=0,width=0.4\textwidth]{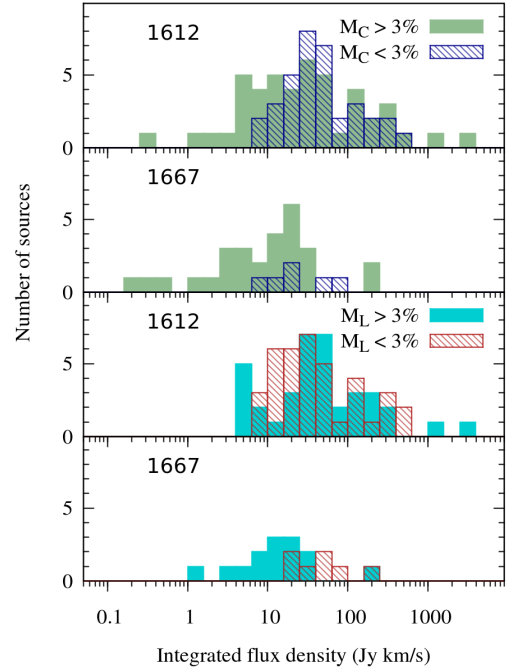}
  \caption{Histograms of  all sources versus the integrated Stokes $I$ flux density for weak and  more strongly
           polarized 1612 and 1667\,MHz lines. The fractional polarizations $M_{\mathrm C}$ and $M_{\mathrm L}$ are
           the ratios of the integrated flux densities.  Note that there are statistically significant differences
($p<0.01$) between the two distributions in each panel. \label{int_fluxes}}
\end{figure}

There are several sources in the sample that exhibit complex profiles in the Stokes $V$ and $p$
so that their polarization properties can only be properly determined using the integrated flux 
densities of the Stokes parameters. The number of sources versus the integrated Stokes $I$ 
flux density for both OH lines are shown in Fig. \ref{int_fluxes}. Here $M_{\mathrm C}=V_{\mathrm {int}}/I_{\mathrm {int}}$
and $M_{\mathrm L}=p_{\mathrm {int}}/I_{\mathrm {int}}$, where $I_{\mathrm {int}}$, $V_{\mathrm {int}}$ and  $p_{\mathrm {int}}$ 
are the integrated Stokes $I$, $V$ and $p$ flux densities, respectively.
 A distinction is made between weakly ($M_{\mathrm C}$,
$M_{\mathrm L}$ $<$3\%) and more strongly ($>$3\%) polarized sources. 
There is evidence for higher circular and linear polarization in the 1667\,MHz sources with
$I_{\mathrm {int}}<$10-20\,Jy\,km\,s$^{-1}$. Less polarized 1667\,MHz sources are usually the brightest
in the sample. For the 1612\,MHz line there is a slight excess of weak ($<$10\,Jy\,km\,s$^{-1}$) 
sources with $M_{\mathrm C}>$3\%. For the majority of the 1612\,MHz sources $M_{\mathrm C}$ and $M_{\mathrm L}$ 
do not depend on the integrated Stokes $I$ flux density.
 Here the defined ratios estimate the net circular and linear polarizations.
The net polarization would be zero for a uniformly filled maser shell with 
symmetric magnetic and velocity fields and density gradient. Hence, the trend 
to higher integrated fractional circular polarization at lower integrated total 
intensity may suggest that these objects have less homogeneous velocity
and/or density distributions.

\subsection{Highly polarized sources}

 Fractional polarizations referred to in this section as $m_{\mathrm C}$ and $m_{\mathrm L}$ are determined
 at the velocities of the $V$ and $p$ peaks.
Table \ref{strongly_pol_tab} lists the sources with fractional circular and linear polarizations
greater than 20\% and/or 10\%, respectively. There are 26 sources with $m_{\mathrm C}>$20\% and 17 sources
with $m_{\mathrm L}>$10\%.

\begin{table}
\caption{Sources with highly polarized OH features. The fractional polarizations for the Stokes $V$ 
         and $p$ peaks, $V_{\mathrm p}$ and $p_{\mathrm p}$ are given. For the Stokes $I$ peak the fractional
         polarization is also given when we were able to reliably determine it. \label{strongly_pol_tab}}
\begin{tabular}{l c c c c c}

\hline

\multicolumn{1}{c}{Name} & Line 

&\multicolumn{1}{c}{ $m_{\mathrm C}(V_{\mathrm p})$ }

&\multicolumn{1}{c}{ $m_{\mathrm C}(I_{\mathrm p})$ }

&\multicolumn{1}{c}{ $m_{\mathrm L}(p_{\mathrm p})$ }

&\multicolumn{1}{c}{ $m_{\mathrm L}(I_{\mathrm p})$ }\\  

&(MHz)&\multicolumn{4}{c}{(\%)}\\

\hline

  &  &  &  &   \\

V CVn            & 1667&   96.4&       &     &      \\

RT Vir           & 1665&   14.9&   10.4& 14.5& 14.5 \\               

                 & 1667&$-$90.7&$-$76.7&     &      \\

R Cnc            & 1665&   83.8&   17.2& 47.8& 41.3 \\

                 & 1667&   90.2&       &     &      \\

Z Pup            & 1665&   89.9&       &     &      \\

                 & 1667&   77.2&       &     &      \\

IRAS17150$-$3754 & 1612&$-$89.8&       &     &      \\

V778 Cyg         & 1665&   58.1&   58.1& 49.1& 34.9 \\

                 & 1667&   83.6&   63.8&     &      \\

IRAS16342$-$3814 & 1612&$-$82.7&$-$82.7& 15.7&  6.5 \\

OH 17.7$-$2.0    & 1612&$-$81.5&$-$81.5& 13.7& 13.7 \\

                 & 1665&$-$67.2&$-$67.2& 11.5& 11.5 \\

                 & 1667&$-$39.8& $-$2.1&  8.9&  8.9 \\

GY Aql           & 1665&   25.5&   12.5& 12.2&  6.2 \\

                 & 1667&$-$50.4&       & 28.4& 16.2 \\

UX Cyg           & 1612&   44.9&       & 31.7&      \\

R Cas            & 1665&   43.9&   39.2&     &      \\

U Her            & 1665&$-$43.6& $-$2.4&  7.6&  3.4 \\

                 & 1667&$-$28.3& $-$5.3&  9.3&  5.6 \\

V524 Cas         & 1612&   42.9&   21.3&     &      \\

R LMi            & 1665&   40.1&   25.3&     &      \\

OH 1.2+1.3       & 1667&$-$35.4&$-$18.1& 15.0&      \\

W Hya            & 1665&$-$35.0&    9.4& 30.4& 19.2 \\

                 & 1667&$-$24.8&$-$10.6& 15.0& 14.3 \\

IRAS17115$-$3322 & 1612&$-$30.5&$-$26.3&     &  7.7 \\

                 & 1667&   33.1&   14.1&     &      \\

OH31.0+0.0       & 1667&   27.9&   27.9&     &      \\

R Crt            & 1665&$-$27.3&$-$22.1& 28.3&      \\

                 & 1667&$-$24.0&$-$17.6& 17.0&      \\

U Ori            & 1667&$-$26.2&       &     &      \\

V1300 Aql        & 1667&$-$26.9&$-$26.9&  4.7&  4.7 \\

OH 31.0$-$0.2    & 1612&   26.1&   26.1& 12.6&  2.1 \\

OH 0.9+1.3       & 1667&$-$24.7&       &     &      \\

OH 12.8$-$1.9    & 1612&$-$21.9&    5.3&  8.5&  8.5 \\

RR Aql           & 1667&$-$21.6& $-$9.7&  7.9&      \\

OH 138.0+7.2     & 1667&$-$20.6&$-$20.6& 19.7&      \\

OH 55.0+0.7      & 1667&       &       & 19.4&  3.3 \\

OH 26.5+0.6      & 1667&       &       & 16.3&  2.8 \\

OH 53.6$-$0.2    & 1667& $-$5.2& $-$5.2& 13.1& 11.8 \\

IW Hya           & 1612&       &       & 11.3&  2.5 \\

OH 25.1$-$0.3    & 1612&       &       & 10.4& 10.4 \\

\hline

\end{tabular}

\end{table}

\begin{figure*}   
\centering
\resizebox{0.65\hsize}{!}{\includegraphics[angle=-90]{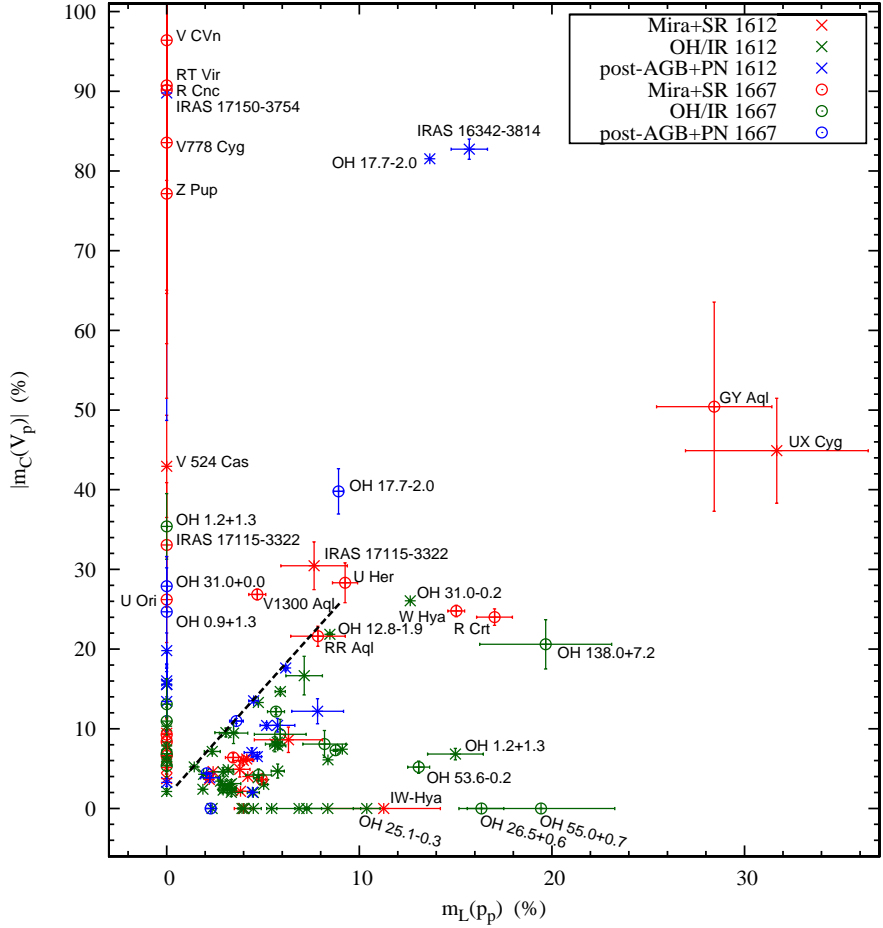}}
\caption{Diagram of the circular ($m_{\mathrm C}$) and linear ($m_{\mathrm L}$) fractional polarizations
         for the whole sample. The fractional polarization $m_{\mathrm C}$ and $m_{\mathrm L}$ are determined
         for the Stokes $V$ and $p$ peaks, respectively. The sources with a polarized flux density
         lower than 3$\sigma$ detection level are assumed to have a fractional polarization equal
         to zero. The points of $|m_{\mathrm C}|>20\%$ or/and $m_{\mathrm L}>10 \%$ are labelled with the names of the
         sources. The thick dashed line defines the relation
         between the fractional polarizations of the form $m_{\mathrm C}=2.7m_{\mathrm L} + 1.5$.\label{strongest_mlmc}} 

\end{figure*}

This subsample contains 10 and 21 objects with the 1612 and 1667\,MHz lines, 
respectively. About two thirds of the 1667\,MHz highly polarized sources are the Mira$+$SR variables.
Fig. \ref{strongest_mlmc} shows the relation between the circular and linear fractional polarizations
for the whole sample. The majority of moderately polarized sources, i.e. with well-determined 
$|m_{\mathrm C}|<$25\% and $m_{\mathrm L}<$10\% lies below the line of form $|m_{\mathrm C}|=2.7m_{\mathrm L}+1.5$.
There are a few groups of outliers; the group of Mira$+$SR objects with strongly ($|m_{\mathrm C}|>$75\%) 
polarized 1667\,MHz features without measurable linear polarization, the two OH/IR objects (OH17.7$-$2.0
and IRAS16342$-$3814) with features of $|m_{\mathrm C}|>$80\% and $m_{\mathrm L}\la$15\%, the two Miras (GY\,Aql
and UX\,Cyg) with features of high linear and circular polarization, the group mainly of OH/IR objects
with features of $m_{\mathrm L}>$10\% and low or marginal circular polarization.
We notice that the sources with highly polarized features are usually the Mira$+$SR objects and a few
OH/IR objects showing outburst activity in OH maser lines 
(e.g. UX\,Cyg - Etoka \& Le Squeren (\cite{etoka00}), OH17.7-2.0 - Szymczak \& Gerard (\cite{szymczak05})).

\subsection{Polarization angle arrangement}

\begin{figure}   
   \includegraphics[angle=0,width=0.5\textwidth]{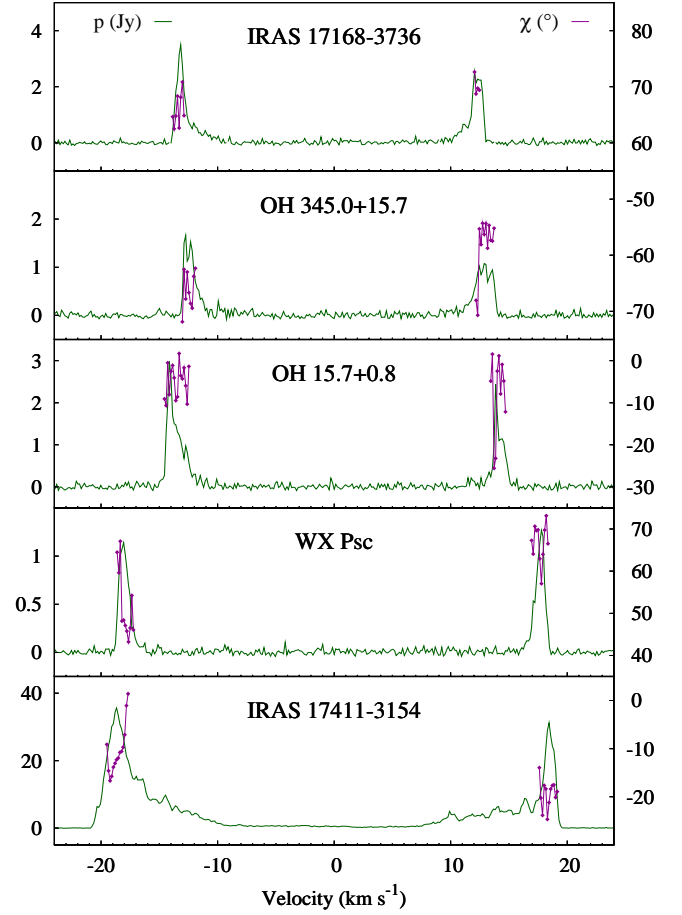}
  \caption{Examples of the 1612\,MHz $p$ spectra in the sources with the emission
           appeared at the blue- and red-shifted velocities. The polarization position
           angles (points) are superimposed (right ordinate) for the extreme 
           outside channels. The velocity is measured with regard to the systemic
           velocity.
\label{angles}}
\end{figure}

There are 29 1612\,MHz sources with linearly polarized emission in both parts of the double peaked
profile. 
 Several exemplary
 $p$ spectra with the superimposed position angle of linear polarization
are shown in Fig. \ref{angles}.
 Most of the sources show only weak ($<$15\degr)
 variations of the polarization angle from channel to channel.  Table
 \ref{delta_chi_tab} gives the average polarization angles for the four most blue- and
 red-shifted channels in the 1612\,MHz $p$ profiles ($\chi_{\mathrm
 B}$, $\chi_{\mathrm R}$) and the difference ($\Delta\chi_{\mathrm
 {BR}}$). We found that $\Delta\chi_{\mathrm {BR}}$ is lower
 than 20\degr   
for two thirds (20/29) of the sources and is only higher than 70\degr \,in two sources. 
For the whole subsample the absolute average and median values of $\Delta\chi_{\mathrm BR}$ are 
19\fdg3$\pm$4.0 and 11\fdg6, respectively. For a few sources observed at several epochs spanning 
6-8 years, no systematic variations of $\Delta\chi_{\mathrm BR}$ higher than 10\degr\, were found.

In the standard circumstellar shell model (Reid et al.\,\cite{reid77}) the 1612\,MHz double peaks
come from two very compact regions of opposite sides of the shell. VLBI observations of the archetypal 
OH/IR object OH127.8$+$0.0 proved that the 1612\,MHz emission of $\sim$0.5\,km\,s$^{-1}$ width 
from the near and far sides of the circumstellar shell is not resolved with a 0\farcs03 beam 
(Norris et al.\,\cite{norris84}). 
 Because the diameter 
of the OH shell is 3\farcs8 (Bowers \& Johnston\,\cite{bowers90}), the size of the compact area unresolved 
with the VLBI is less than 0.8\% of the shell size. Therefore it is very likely that the emission from
both extreme sides of the shell is not depolarized if observed with the NRT and it reliably probes 
a local magnetic field. Small differences in the polarization angles at the near and far edges 
of the OH shells in our subsample suggest a regular magnetic field geometry in the 1612\,MHz maser 
regions. 
However, it is hard to explain this homogeneous magnetic field over OH shells of large diameters 
(e.g. Habing\,\cite{habing96}). A plausible explanation for the low values of $\Delta\chi_{\mathrm BR}$  
is that the origin of the field is intrinsic and the same field orientation is carried away (frozen in) 
by the stellar wind flowing towards the front and back of the envelope. Alternatively, the magnetic field 
could be of galactic origin and be amplified and/or distorted by the stellar wind.
 High values of $\Delta\chi_{\mathrm BR}$ in a minority of sources suggest a deviation from globally ordered
magnetic field caused by local outflows or outbursts. The latter possibility is likely for OH17.7$-$2.0 where
$\Delta\chi_{\mathrm BR}$ is only $-$3\fdg7 (Tab. \ref{delta_chi_tab}) but the polarization angle of the 
eruptive feature near 73\,km\,s$^{-1}$ (Szymczak \& G\'erard\,\cite{szymczak05}) is about $-$15\degr\,
and differs by about 64\degr\, from the mean value for the extreme blue- and red-shifted velocities
(Tab. \ref{delta_chi_tab}).

Nine sources in the sample have linearly polarized emission at the two frequencies from the same
usually red-shifted side of the shell (Tab. A.1, Fig. B.1). 
 The scatters of $\chi$ angles within the OH mainline channels are on average 2.7 times larger than 
those within 1612\,MHz lines. This may suggest that the 1667\, MHz masers come from more
 turbulent regions and/or the magnetic fields are less ordered.  
For seven sources
the velocities of the extreme emission at both frequencies overlap within less than 0.1\,km\,s$^{-1}$.
In the two objects OH127.8$-$0.0 and OH53.6$-$0.2 the difference in the average polarization angle for 
four neighbouring channels of the 1612 and 1667\,MHz lines is less than 18\degr~ while for the remaining 
sources it ranges from 36 to 80\degr. The mean difference for nine sources is 44$\pm$8\degr~ 
and the median is 47\degr. This result suggests that the magnetic fields probed by the two lines 
are generally not aligned.

\begin{table}

\caption{Averaged polarization position angles $\chi_{\mathrm B}$ and $\chi_{\mathrm R}$ for the extreme
         blue- and red-shifted parts of the 1612\,MHz spectrum, respectively. In most cases
         the number of averaged channels is four. 
         The $\Delta\chi_{\mathrm BR}$ is a difference between $\chi_{\mathrm B}$ and $\chi_{\mathrm R}$.
         The standard deviations (SD) are given in the parentheses. 
         \label{delta_chi_tab}}

\begin{tabular} { l r r r}

\hline

     Name   & $\chi_{\mathrm B}$ (SD)  & $\chi_{\mathrm R}$ (SD) &  $\Delta\chi_{BR}$ (SD) \\        

\hline

WX Psc            &   52.7(4.8)     &   63.4(2.6)    &  $-$10.8(5.4) \\

OH127.8$+$0.0     &$-$20.5(4.4)     &$-$33.6(3.3)    &     13.1(5.4) \\

OH138.0$+$7.2     &$-$47.3(2.5)     &$-$34.2(2.3)    &  $-$13.1(3.4) \\

IK Tau            &   15.8(2.1)$^3$ &   30.7(5.2)$^3$ &  $-$14.8(5.6) \\

OH345.0$+$15.7    &$-$65.0(1.3)     &$-$56.1(0.7)    &   $-$8.8(1.5) \\

IRAS17168$-$3736  &   66.6(1.8)     &   68.2(1.7)    &   $-$1.6(2.5) \\

IRAS17177$-$3627  &   32.7(1.6)     & $-$9.2(1.7)    &     41.8(2.3) \\

IRAS17271$-$3425  &   80.6(3.2)     &   74.1(2.7)    &      6.5(4.2) \\

OH1.2+1.3         &   79.3(1.0)$^3$ &   80.3(2.5)    &   $-$1.0(2.7) \\

IRAS17411$-$3154  &$-$13.7(0.7)     &$-$18.7(0.8)    &      4.9(1.1) \\

OH1.1$-$0.8       &$-$59.8(2.3)     &$-$83.9(6.4)    &     24.1(6.8) \\

IRAS17579$-$3121  &$-$74.2(6.3)     &$-$66.1(8.1)    &   $-$8.1(10.3)\\

OH15.4$+$1.9      &$-$74.3(4.3)     &$-$89.3(3.9)    &     15.0(5.8) \\

OH15.7$+$0.8      & $-$4.0(1.5)     & $-$3.1(2.0)    &   $-$0.8(2.5) \\

OH12.8$-$1.9      &$-$19.5(2.3)     & $-$9.4(1.1)    &  $-$10.2(2.5) \\

OH16.1$-$0.3      &   76.5(3.5)     &   90.6(2.4)    &  $-$14.1(4.2) \\

OH17.7$-$2.0      &$-$76.8(6.0)$^3$ &$-$80.4(7.8)    &  $-$3.7(9.8) \\

OH25.1$-$0.3      &   44.0(4.8)     &   72.6(6.7)    &  $-$28.6(8.2)\\

IRAS18445$-$0238  &$-$54.3(4.0)     &$-$46.2(2.7)    &    $-$8.1(4.8)\\

OH32.8$-$0.3      &   15.6(3.2)     &$-$15.4(10.3)$^3$&    31.0(10.8)\\

OH39.7$+$1.5      &$-$57.5(3.3)     &$-$73.0(3.4)$^3$&     15.5(4.7) \\

OH37.1$-$0.8      & $-$2.3(3.1)     &$-$13.9(19.2)$^2$&   11.6(19.4) \\

IRAS19059$-$2219  &$-$47.3(3.1)     &$-$44.2(7.3)    &   $-$3.1(7.9)\\

OH42.3$-$0.1      &   38.8(21.9)$^3$&$-$58.3(4.1)    &  $-$82.9(22.3)\\

OH45.5$+$0.0      &$-$32.4(7.8)     &   17.4(10.1)   &  $-$49.8(12.8)\\

OH55.0$+$0.7      &   24.8(1.4)     &   20.9(5.4)    &      3.9(5.6) \\

OH77.9$+$0.2      & $-$4.0(4.0)     &   67.5(7.5)$^3$&  $-$71.4(8.5)\\

OH75.3$-$1.8      &    2.4(2.6)     &   13.1(1.6)   &   $-$10.7(3.0) \\

OH104.9$+$2.4     &  107.7(7.6)     &   78.0(2.6)    &     29.7(8.0)\\

\hline

\end{tabular}

$^3$ three, $^2$ two averaged channels

\end{table}

\subsection{Possible Zeeman pairs}

\begin{figure}   
   \includegraphics[angle=-90,width=0.45\textwidth]{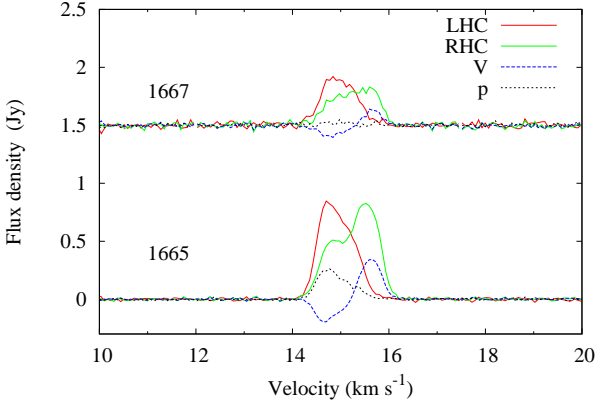}
  \caption{Tentative Zeeman splitting in the OH maser lines of R\,Cnc. The left- (LHC) 
           and right-hand (RHC) circular polarization spectra, $V$ Stokes and $p$ spectra
           are shown for the two frequencies. \label{rcnc-zeeman}}
\end{figure}

\begin{figure} 

\includegraphics[angle=-90,width=0.45\textwidth]{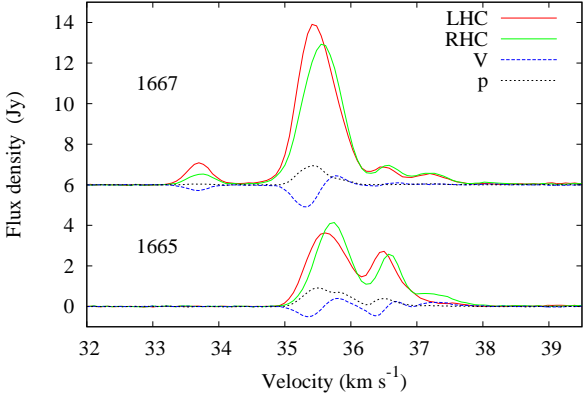}

  \caption{Same as in Fig. \ref{rcnc-zeeman} but for W\,Hya. \label{whya-zeeman}} 

\end{figure}

The main line $V$ Stokes spectra of R\,Cnc and W\,Hya contain S-shaped features possibly 
 caused by the Zeeman effect (Figs. \ref{rcnc-zeeman} and \ref{whya-zeeman}). Although this
assumption needs to be verified by VLBI observations one, can roughly estimate the strength of the 
magnetic field assuming that the ratio of the Zeeman splitting to the Doppler linewidth is 
higher or equal to unity. In this regime the separation between the peaks of $V$ profile
is a direct measure of the field strength (Elitzur\,\cite{elitzur96}).
We used the Zeeman splitting coefficient of 0.5886\,km\,s$^{-1}$\,mG$^{-1}$ and
0.3524\,km\,s$^{-1}$\,mG$^{-1}$ for the 1665 and 1667\,MHz lines, respectively. For the 1665 and
1667\,MHz features of R\,Cnc near 15.2\,km\,s$^{-1}$ the field strength estimates along the line 
of sight are $-1.31\pm$0.07\,mG and $-2.26\pm$0.11\,mG, respectively. For the strongest feature 
of W\,Hya near 35.6\,km\,s$^{-1}$ the field strength is $-0.32\pm$0.03\,mG and $-0.36\pm$0.04\,mG 
at the 1665 and 1667\,MHz lines, respectively. A negative sign means a field pointed towards the observer.
 These estimates should be treated with caution because the Zeeman patterns are far from  
perfect. Nevertheless, in
both sources the field estimates are quite similar at both frequencies and moreover
they are consistent with the high angular resolution measurements of other AGB and post-AGB stars
(e.g. Szymczak et al.\,\cite{szymczak98}; Bains et al.\,\cite{bains03, bains09}). The presence of 
polarized features in lots of our sources suggests that many Zeeman pairs will be found in 
high angular resolution studies.

\subsection{OH 1667\,MHz overshoot}

There are 63 objects in the sample for which the spectral extent of 1667\,MHz emission is higher 
by at least 0.2\,km\,s$^{-1}$ than that of 1612\,MHz emission. This overshoot effect is seen
either at the blue- or red-shifted or at both edges. A summary of the mean and median expansion
velocities and overshoots for the three classes of objects is given in Table \ref{overshoots_tab}. 
Here, the expansion velocity is determined as a half of the peak-to-peak velocity width of the
1612\,MHz profile. The class of Mira$+$SR stars has expansion velocities significantly 
lower than in the two other classes. The lowest mean and median overshoots of 
0.78$\pm$0.16\,km\,s$^{-1}$ and 0.47\,km\,s$^{-1}$, respectively, belong to the Miras 
and SR objects. The highest median value of overshoot of 2.08\,km\,s$^{-1}$) is determined
for the post-AGB$+$PN class. The phenomenon is most frequent (63\%) in the OH/IR objects.

The overshoot effect was broadly discussed by Sivagnanam \& David (\cite{siva99}) for a sample
of 21 AGB and post-AGB objects who proposed several mechanisms. Deacon et al. (\cite{deacon04})
suggested the existence of some acceleration in the outer parts of the shell where the 1612 and 1667\,MHz
maser originate and that the 1667\,MHz masers are located farther than the 1612\,MHz ones.
The trend in overshoot found in our sample suggests that the phenomenon can be related to 
the evolutionary sequence. We did not find any obvious relationships between overshoot and the polarization 
parameters.

\begin{table*}

\caption{Average and median values of expansion velocities and overshoots for the three classes of sources.
\label{overshoots_tab}}

\begin{tabular} {c c c c c c c c}

\hline

Class & Line  & Number & \multicolumn{2}{c}{Expansion velocity (km\,s$^{-1}$)} & Number & 

\multicolumn{2}{c}{Overshoot (km\,s$^{-1}$)} \\ 

 \cline{4-5}

 \cline{7-8}

     & (MHz) &        & Mean(SD)   & Median &    & Mean(SD)    & Median \\ 

\hline

Mira+SR & 1612 & 23 & 10.68(1.08) & 12.16 &     &            &      \\

         & 1667 & 26 &  8.36(0.98) &  6.12 & 16  & 0.78(0.16) & 0.47 \\

OH/IR    & 1612 & 56 & 15.15(0.57) & 14.38 &     &            &      \\

         & 1667 & 37 & 15.36(0.54) & 15.09 & 37  & 1.18(0.14) & 0.93 \\

post-AGB+PN & 1612 & 14 & 12.08(1.14) & 12.36 &     &            &      \\

         & 1667 & 8  & 13.34(0.74) & 13.51 & 10  & 2.25(0.62) & 2.08 \\

\hline

\end{tabular}

\end{table*}

\section{Discussion}
\subsection{Occurence of polarization}
The present full-polarization data for a substantial sample of AGB and post-AGB objects have allowed 
us to give a comprehensive picture of the polarization properties of OH circumstellar masers. 
First of all, our high sensitivity and spectral resolution study shows that the polarized OH maser 
emission is widespread. This finding is at odds with early reports, which demonstrated a paucity of 
polarized features in the 1612\,MHz spectra and only occasional detections of linear polarization 
at 1667\,MHz (Cohen\,\cite{cohen89}, for review). Thus, spectral resolution appears to be
a crucial factor (Cohen et al.\,\cite{cohen87}; Zell \& Fix\,\cite{zell91}).

 The study allowed us to extract a complete subsample of 1612\,MHz emitters, composed of
8 Miras+SR stars, 38 OH/IR objects and 10 post-AGB+PN sources, which is not biased by the sensitivity 
limit and the minimum believable fractional polarization (see Sect. 3.2 for definition). In this
subsample the detection rate of polarized features is 75-90\% and there is no intrinsic difference
in the occurrence of polarized emission between the three classes of studied objects. 
A complete subsample of 1667\,MHz sources contains only 2 Miras+SRs, 3 OH/IR objects and 4 post-AGB+PN 
stars. All but one of these targets show polarized emission. It seems that the occurrence of polarized
features at 1667\,MHz does not differ between the three classes of stars.

\subsection{Effect of depolarization}

For 1612\,MHz sources the mean and median degrees of circular polarization in the post-AGB+PN objects
are significantly higher that those in the two other classes of studied stars. We notice that
in the sample are at least two bursting post-AGB stars; IRAS16342-3814 and OH17.7-2.0 with $m_{\rm C}>$80\%.
These objects sometimes experience OH bursts of highly polarized emission at 1612\,MHz usually not 
associated with bursts at 1665 and 1667\,MHz (Szymczak \& G\'erard\,\cite{szymczak05}).      
In many post-AGB objects there is a disruption of detached AGB 1612\,MHz shells that are no longer 
spherically symmetric and the appearance of a well-defined bipolar morphology at 1667\,MHz is observed
(Zijlstra et al.\,\cite{zijlstra01}). 
 In these partially-filled 1612-MHz structures, observed at high angular
resolution, Zeeman components are much less spatial blended than in dense 
($10^{-4}$M$_{\sun}$, Habing\,\cite{habing96}),
more fully filled shells that are observed around OH/IR objects 
(e.g. Bowers \& Johnston\,\cite{bowers90}).
 Therefore, the enhancement of the circular
polarization degree at 1612\,MHz in the post-AGB+PN objects compared to the two other classes of objects
 could be caused by negligible depolarization in the first class of objects.
However, the post-AGB+PN group may be heterogeneous because of the lack of accurate criteria 
for selection (Sect. 2.1) and significant depolarization cannot be excluded in some objects 
(Szymczak \& G\'erard\,\cite{szymczak04}).

The mean degree of circular polarization at 1667\,MHz is significantly higher in the Mira+SR stars
than in the other classes of objects. The mass loss rates for Mira and SR variables are about 
$10^{-7}$M$_{\sun}$ (Habing\,\cite{habing96}) while the 1667\,MHz maser originates in the inner parts 
of the shell at a distance of about 50-150\,AU from the central star (Chapman et al.\,\cite{chapman94}; 
Szymczak et al.\,\cite{szymczak98}). This implies that regardless of the magnetic field type, its strength 
in the 1667\,MHz maser regions should be higher than in the 1612\,MHz OH maser outer shells. 
High angular resolution images revealed a highly fragmentary distribution of the 1667\,MHz emission 
(Chapman et al.\,\cite{chapman94}; Szymczak et al.\,\cite{szymczak98}), indicating a low filling factor 
for the OH molecule. In this shell the spatial overlapping of Zeeman components is weak, 
if there is any at all. 
Furthermore, Miras and SR variables often exhibit erratic variability of OH emission, which usually 
is highly polarized (Etoka \& Le Squeren\,\cite{etoka97}). We conclude that the higher degree of circular 
polarization in the Mira+SR stars compared to the other classes of objects is because of negligible
depolarization in the envelopes that are partially filled by the gas 
 clouds sustaining OH maser emission.

\subsection{Zeeman splitting}
The interpretation of OH maser polarization data strongly depends on the assumed value of the ratio of Zeeman
splitting to Doppler linewidth, $x_{\mathrm B}$, (Elitzur\,\cite{elitzur96}). For $x_{\mathrm B}>1$ the separation of 
the $\sigma$ components of the Zeeman pair scales directly to the strength of the magnetic field along the line
of sight. If $x_{\mathrm B}<1$ only the Doppler width can be estimated from the separation of the two peaks of
$V$ Stokes profile. High angular resolution observations of OH lines in some AGB and post-AGB stars 
(Szymczak et al.\,\cite{szymczak98}; Bains et al.\,\cite{bains03, bains04, bains09}; Amiri et al.\,\cite{amiri10})
revealed a magnetic field strength of 1$-$3\,mG. This value agrees well with our tentative estimates 
of magnetic fields of 0.3$-$2.3\,mG in two Miras. The kinematic temperature in regions of circumstellar
1612 and 1667\,MHz masers is 50-100\,K (Goldreich \& Scoville\,\cite{goldreich76}; Fong et al.\,\cite{fong02}),
hence for OH molecule the thermal linewidth is 0.36$-$0.51\,km\,s$^{-1}$. It implies that $x_{\mathrm B}>1$
is fulfilled for most sources in the sample. 
 We note that the Zeeman pairs are of unequal strength, and in several cases likely only one sense of circularly
polarized emission is seen. This can occur when the maser radiation is amplified in the region where
the gradients of the velocity match the magnetic field (Cook\,\cite{cook66}). 
 A simpler explanation for uneven circular polarization is 
the overlap of different Zeeman components caused by a velocity gradient within the maser region
(Deguchi \& Watson\,\cite{deguchi86}).

 In the case of $x_{\mathrm B}>1$
the polarization of the $\sigma$
components is described by the equations $V/I$=2cos$\theta$/(1+cos$^2\theta)$ 
and $Q/I$=sin$^2\theta$/(1+cos$^2\theta)$ (Goldreich et al.\,\cite{goldreich73}; Elitzur\,\cite{elitzur96}),
where $\theta$ is the angle between the magnetic field direction and the line of sight.
 We used these equations to model the $m_{\mathrm C} - m_{\mathrm L}$ diagram. It is easy to show that 
the ratio $V/I$ integrated and averaged over the solid angle builds up rapidly between 
$\theta=0$\degr\, and $50$\degr\, while the $Q/I$ has barely started growing. This means that 
the ratio $V/Q$ is very high until $45$\degr\, and that may explain the {\it void} in Fig. \ref{strongest_mlmc} 
even with a spherically symmetric distribution of the $\theta$ angle. 
Although our study was of insufficient angular resolution to determine polarization properties of individual
maser spots and the strongest polarized features were used to construct the $m_{\mathrm C} - m_{\mathrm L}$ diagram, 
the appearance of the {\it void} (Fig. \ref{strongest_mlmc}) may be the proof that the polarized features 
are the $\sigma$ components and that the condition $x_{\mathrm B}>1$ is fulfilled for unsaturated OH masers.

 Fig. \ref{strongest_mlmc} shows a distinct group of 1612/1667 Mira+SR sources 
with only circularly polarized emission (the first six objects in Tab. \ref{strongly_pol_tab}), 
indicating that the magnetic field is 
parallel to the line of sight. Another small group of mainly OH/IR objects (the five last objects in 
Tab. \ref{strongly_pol_tab}) with predominant linear polarization at 1612 or 1667\,MHz represents 
the case where the magnetic field is perpendicular to the line of sight. These objects appear
  to be good candidates for high angular resolution studies to determine their 
properties better and to examine if significant differences in polarization parameters could be caused by 
different states of saturation as predicted in the standard model of maser polarization 
(Goldreich et al.\,\cite{goldreich73}; Vlemmings\,\cite{vlemmings06}).

\section{Conclusions}

We have presented the results of full-polarization observations of the 1612 and 1667\,MHz OH maser 
lines towards a sample of 117 AGB and post-AGB stars. A complete set of full-polarization spectra 
and the basic polarization parameters was given.

 The polarized features occur in more than 75\% of the sources in the sample complete to the total 
flux density higher than 7.5\,Jy and the minimum fractional polarization of 2\%. There is no intrinsic
difference in the occurrence of polarized emission between the three classes of objects. 
The highest fractional polarization occurs for the post-AGB+PN and the Mira+SR classes at 1612 and 
1667\,MHz, respectively. Differences in the fractional polarizations are likely caused by depolarization 
 caused by blending of polarized emission.

 The relationship between the circular and linear fractional polarizations for extreme red- and 
blue-shifted features, which are less likely to be blended, was interpreted within the frame of the standard 
models of polarization for the regime of the Zeeman splitting higher than the Doppler line width. 
The strength of the magnetic field derived from the tentative Zeeman pairs for a few Miras of 0.3-2.3\,mG
is consistent with the results of high angular resolution observations of other AGB stars. 
 The front and back caps show only small differences in polarization angle ($<$20\degr) at 1612\,MHz,
which suggests a regular geometry of the magnetic field in the outer parts of the shells. 
 The observed relation between the circular and linear fractional polarizations 
suggests that the polarized features are the $\sigma$ components and the 
Zeeman splitting is higher than the Doppler line width.

The study has provided many targets for future investigations. 
 VLBI\,+\,e-MERLIN mapping 
of polarized OH maser emission will be essential to measure the magnetic field strength and its structure and 
to refine the present results.

\begin{acknowledgements}

P.W. acknowledges support by the UE PhD Scholarship Programme (ZPORR).
The Nancay Radio Observatory is the Unit\'e Scientifique de Nancay of 
the Observatoire de Paris, associated with the CNRS. The Nancay Observatory
acknowledges the financial support of the Region Centre in France.

\end{acknowledgements}


\begin{thebibliography}{}

\bibitem[2010]{amiri10} Amiri, N., Vlemmings, W., \& van Langevelde, H.J., 2010, \aap, 509, 26
\bibitem[2003]{bains03} Bains, I., Gledhill, T.M., Yates, J.A., \& Richards, A.M.S. 2003, \mnras, 338, 287
\bibitem[2004]{bains04} Bains, I., Gledhill, T.M., Richards, A.M.S., \& Yates, J.A. 2004, \mnras, 354, 529
\bibitem[2009]{bains09} Bains, I., Richards, A.M.S., \& Szymczak, M. 2009, ASPC, 404, 368
\bibitem[1979a]{baud79a} Baud, B., Habing, H.J., Matthews, H.E., \& Winnberg, A. 1979a, \aaps, 35, 179
\bibitem[1979b]{baud79b} Baud, B., Habing, H.J., Matthews, H.E., \& Winnberg, A. 1979b, \aaps, 36, 193

\bibitem[1979]{benson79} Benson, J.M., Mutel, R.L., Fix, J.D., \& Claussen, M.J. 1979, \apj, 229, 87

\bibitem[1990]{benson90} Benson, P.J., Little-Marenin, I.R., Woods, T.C., et al. 1990, \apjs, 74, 911

\bibitem[1987]{bedijn87} Bedijn, P. 1987, \aap, 186, 136

\bibitem[1990]{bowers90} Bowers, P.F., \& Johnston, K.J. 1990, \apj, 354, 676

\bibitem[1978]{bowers78} Bowers, P.F., 1978, \aaps, 31, 127

\bibitem[1986]{chapman86} Chapman, J.M., \& Cohen. R.J. 1986, \mnras, 220, 513

\bibitem[1994]{chapman94} Chapman, J.M., Sivagnanam, P., Cohen, R.J., \& Le Squeren, A.M. 1994, \mnras, 268, 475

\bibitem[1982]{claussen82} Claussen, M.J., \& Fix, J.D. 1982, \apj, 263, 153

\bibitem[1989]{cohen89} Cohen, R.J. 1989, RPPh, 52, 881

\bibitem[1987]{cohen87} Cohen, R.J., Downs, G., Emerson, R., et al. 1987, \mnras, 225, 491

\bibitem[1966]{cook66} Cook, A.H. 1996, Nature, 211, 503

\bibitem[1986]{deguchi86} Deguchi, S., \& Watson, W.D. 1986, \apj, 300, L15

\bibitem[2004]{deacon04} Deacon, R.M., Chapman, J.M., \& Green, A.J. 2004, \apjs, 155, 595 

\bibitem[1988]{eder88} Eder, J., Lewis, B.M., \& Terzian, Y. 1988, \apjs, 66, 183

\bibitem[1996]{elitzur96} Elitzur, M. 1996, \apj, 457, 415

\bibitem[2008]{engels08} Engels, D., \& Bunzel, F. 2008, Database of Circumstellar Masers v1.2  
  (http://www.hs.uni-hamburg.de/maserdb) 

 \bibitem[1997]{etoka97} Etoka, S., \& Le Squeren, A.M. 1997, \aap, 321, 877

\bibitem[2000]{etoka00} Etoka, S., \& Le Squeren, A.M. 2000, \aaps, 146, 179

\bibitem[1979]{fix79} Fix, J. D. 1979, \apj, 232, 39

\bibitem[2002]{fong02} Fong, D., Justtanont, K., Meixner, M., \& Campbell, M.T. 2002, 396, 581

\bibitem[1996]{gray95} Gray, M.D., \& Field, D. 1995, \aap, 298, 243

\bibitem[1973]{goldreich73} Goldreich, P., Keeley, D., \& Kwan, J.Y. 1973, \apj, 179, 111

\bibitem[1976]{goldreich76} Goldreich, P., Scoville, N.Z. 1976, \apj, 205, 144

\bibitem[1996]{habing96} Habing, H.J. 1996, \aapr, 7, 97
 
\bibitem[1981]{jewell81} Jewell, P.R., Webber, J.C., \& Snyder, L.E. 1981, \apj, 249, 118

\bibitem[1991]{lesqueren91} Le Squeren, A.M., Sivagnanam, P., Dennefeld, M., \& David, P. 1991, \aap, 254, 133 

\bibitem[1989]{likkel89} Likkel, L. 1989, \apj, 344, 350

\bibitem[1984]{norris84} Norris, R.P., Booth, R.S., Diamond, P.J., et al. 1984, \mnras, 208, 435

\bibitem[1980]{olnon80} Olnon, F. M., Winnberg, A., Matthews, H. E., \& Schultz, G. V. 1980, \aaps, 42, 1190

\bibitem[1977]{reid77} Reid, M.J., Muhleman, D.O., Moran, J.M., Johnston, K.J., \& Schwartz, P.R. 1977, \apj, 214, 60

\bibitem[1997a]{sevenster97a} Sevenster, M.N., Chapman, J.M., Habing, H.J., Killeen, N.E.B., \& 
  Lindqvist, M. 1997a, \aaps, 122, 79

\bibitem[1997b]{sevenster97b} Sevenster, M.N., Chapman, J.M., Habing, H.J., Killeen, N.E.B., \& 
  Lindqvist, M. 1997b, \aaps, 124, 509

\bibitem[2001]{sevenster01} Sevenster, M.N., van Langevelde, H.J., Moody, R.A., et al. 2001, \aap, 366, 481

\bibitem[1999]{siva99} Sivagnanam, P., \& David, P. 1999, \mnras, 304, 622

\bibitem[1998]{szymczak98} Szymczak, M., Cohen, R.J., \& Richards, A.M.S. 1998, \mnras, 297, 1151

\bibitem[2001]{szymczak01} Szymczak, M., Cohen, R.J., \& Richards, A.M.S. 2001, \aap, 371, 1012

\bibitem[2004]{szymczak04} Szymczak, M., \& G\'erard, E. 2004, \aap,  423, 209

\bibitem[2005]{szymczak05} Szymczak, M., \& G\'erard, E. 2005, \aap,  433, L29

\bibitem[1991]{telintel91} te Lintel Hekkert, P. Caswell, J.L., Habing, H.J., Haynes, R.F., \& 
  Norris, R.P. 1991, \aaps, 90, 327

\bibitem[1988]{vanderveen88} van der Veen, W.E.C.J., \& Habing, H.J. 1988, \aap, 194, 125

\bibitem[1996]{vandriel96} van Driel W., Pezzani J.,  G\'erard E. 1996,
  in High Sensitivity Radio Astronomy, ed. N. Jackson, \& R. J. Davis (Cambridge Univ. Press), 229


\bibitem[2006]{vlemmings06} Vlemmings, W.H.T. 2006, \aap, 445, 1031

\bibitem[1994]{whitelock94} Whitelock, P., Menzies, J., Feast, M., et al. 1994, \mnras, 267, 711

\bibitem[1970]{wilson70} Wilson, W.J., Barrett, A.H., \& Moran, J.M. 1970, \apj, 160, 545

\bibitem[1972]{wilson72} Wilson, W.J., \& Barrett, A.H. 1972, \apj, 17, 385

\bibitem[1972]{wilsonetal72} Wilson, W.J., Schwartz, P.R., Neugebauer, G., Harvey, P.M., \& 
  Becklin, E.E. 1972, \apj, 177, 523

\bibitem[2001]{zijlstra01} Zijlstra, A.A., Chapman, J.M., te Lintel Hekkert, P., et al. 2001, \mnras, 322, 280 

\bibitem[1991]{zell91} Zell, P.J.,\& Fix, J.D. 1991, \apj, 369, 506

\end{thebibliography}
\end{document}